\documentclass[prd,superscriptaddress,twocolumn,
floatfix,preprintnumbers,altaffilletter]{revtex4}
\usepackage{graphicx,url,amssymb,amsmath,longtable,rotating,color,units,
wasysym,subfigure,epsfig,multirow}

\begin{document}

\title{Detecting very long-lived gravitational-wave transients lasting hours to weeks}

\author{Eric~Thrane}
\email{eric.thrane@monash.edu}
\affiliation{LIGO Laboratory, California Institute of Technology, MS 100-36,
Pasadena, CA, 91125, USA}
\affiliation{School of Physics and Astronomy, Monash University, Clayton, Victoria 3800, Australia}

\author{Vuk~Mandic}
\affiliation{School of Physics and Astronomy, University of Minnesota, Minneapolis, Minnesota 55455, USA}

\author{Nelson~Christensen}
\affiliation{Physics and Astronomy, Carleton College, Northfield, Minnesota 55057, USA}

\begin{abstract}
  We explore the possibility of very long-lived gravitational-wave transients (and detector artifacts) lasting hours to weeks.
  Such very long signals are both interesting in their own right and as a potential source of systematic error in searches for persistent signals, e.g., from a stochastic gravitational-wave background.
  We review possible mechanisms for emission on these time scales and discuss computational challenges associated with their detection: namely, the substantial volume of data involved in a search for very long transients can require vast computer memory and processing time.
  These computational difficulties can be addressed through a form of data compression known as coarse-graining, in which information about narrow frequency bins is discarded in order to reduce the computational requirements of a search.
  Using data compression, we demonstrate an efficient radiometer (cross-correlation) algorithm for the detection of very long transients.
  In the process, we identify features of a very long transient search (related to the rotation of the Earth) that make it more complicated than a search for shorter transient signals.
  We implement suitable solutions.
\end{abstract}

\maketitle

\section{Introduction}
Previous work has explored the astrophysics of (and associated detection strategies for) long-lived gravitational-wave transients lasting $\approx$10--$\unit[1000]{s}$; see, e.g.,~\cite{stamp,piro:07,vanputten:08,kiuchi,raffai,murphy_sgr,corsi,pirothrane12,piro:11,stochtrack,stochsky} and references therein.
Most long-transient models rely on the emission of gravitational waves from rotational instabilities in a protoneutron star or its accretion disk but other models are possible as well~\cite{stamp}.
Initial LIGO/Virgo have performed a search for unmodeled long lived transients associated with long gamma ray bursts yielding upper limits on gravitational-wave fluence~\cite{lgrb}.

Despite this progress, there is still a largely unexplored region of parameter space of ``very long-lived'' transient signals lasting hours to weeks; (although, see~\cite{tcw}).
As we discuss below, it may be possible for neutron stars to emit transient gravitational waves on these time scales.
Moreover, exotic models allow for the possibility of a seemingly persistent signal to start or stop during an observing run~\cite{Arvanitaki}, also potentially leading to very long transient signals.
Thus, there is astrophysical motivation to carry out a search for very long lived gravitational wave transients.
An efficient very long transient detection algorithm will have other useful applications: it can establish if an apparently persistent source, e.g., observed in a stochastic background search~\cite{s6vsr23_iso}, exhibits variability in time, and it can be used to understand the behavior of detector artifacts on timescales of days to weeks.

Our goal here is to outline a computationally feasible search for very long-lived transients employing radiometry~\cite{radio_method}.
We review the principles of radiometry below in Section~\ref{radiometry}, but for now, we note that radiometry relies on the cross-correlation of two or more detectors in order to detect gravitational waves as excess coherence.
Excess coherence can be identified through various pattern recognition techniques.
For demonstrative purposes, we here employ a ``seedless clustering'' algorithm~\cite{stochtrack,stochsky}, discussed in greater detail in Section~\ref{radiometry}, though, other pattern recognition techniques may be used as well.

In order to carry out such a search, it is necessary to first address computational challenges that appear, at first glance, to make the search daunting.
Namely, difficulties arise due to the fact that only so much uncompressed spectrographic strain data can be stored at once in random access memory (RAM).
Without compression, it is impossible to hold many hours of data in memory at once on a typical computer.
In theory, data can be read in as needed, but this creates an input/output bottleneck.
Also, in searches that rely on seedless clustering~\cite{stochtrack,stochsky}, the computation time required to sum hours of uncompressed spectrographic data can become prohibitive.

The solution we propose is an extra pre-processing step in which data are compressed in order to facilitate computationally feasible analyses.
Through a procedure known as coarse-graining (already used in searches for the stochastic background~\cite{stoch_s1}), the volume of data can be reduced by orders of magnitude so that days of data can be represented with a spectrogram comparable to one used to represent minutes of uncompressed data.
Coarse-graining necessarily discards information, but through this procedure, we may study much longer transient signals than otherwise possible.

The remainder of this paper is organized as follows.
In Section~\ref{motivation}, we discuss the motivation for searches for very long transients.
In Section~\ref{radiometry}, we review key principles of radiometry needed for subsequent discussion.
(Additional details are included in the Appendix.)
In Section~\ref{compression}, we describe compression procedures for coarse-graining, and in Section~\ref{demonstration}, we demonstrate the recovery of very long-lived signals using compressed data.
In Section~\ref{conclusions}, we summarize our results and suggest some lines of future research.

\section{Motivation}\label{motivation}
The scientific rationale for a search for very-long transients, spanning hours to weeks, was first explored in~\cite{tcw}.
The authors of~\cite{tcw} review a number of somewhat speculative scenarios associated with neutron stars including gravitational-wave emission lasting days to months from non-axisymmetric Ekman flow following a glitch~\cite{glitches1,glitches2,glitches3}, Alfv\'en oscillations from giant magnetic flares (also lasting days to months)~\cite{alfven1,alfven2}, emission from free precession (with a damping time possibly lasting from weeks to years)~\cite{precession1,precession2,precession3}, magnetic instabilities in newborn neutron stars (lasting days)~\cite{maginst}, and gravitational-waves from $r$-modes~\cite{rmode1,rmode2}.

Generic rotational instabilities in newborn neutron stars, potentially powered by fallback accretion~\cite{pirothrane12,melatos1}, may persist on a timescale of hours~\cite{corsi}.
Somewhat more speculatively, we note that observations of intermittent pulsars, which become quiescent on timescales of days (e.g.,~\cite{J1841-0500}), suggest that neutron star dynamics vary on the timescales considered here, motivating exploration of this region of parameter space.
Similarly, variability in accretion on these timescales may affect gravitational-wave emission from accretion-supported mountains~\cite{bildsten}.
Finally, it is worthwhile to be prepared for a surprise: a very long lived transient signal from an unexpected source.
Recent work proposing gravitational-wave emission from gravitationally bound axion clouds~\cite{Arvanitaki}, potentially starting and stopping on the timescale of a few years, serves to illustrate this possibility.
The method we propose below is also applicable to quasi-infinite signals turning on or off during an observing run as well as repeating sources with long-lasting emission periods.

An algorithm for identification of very long transients also provides a means of understanding the time dependence of apparently persistent signals, e.g., in a stochastic background search.
In the event of an ostensibly persistent gravitational-wave detection, it is prudent to investigate if the temporal behavior of the signal is consistent with the assumed signal model.
Matched filter searches for rotating neutron stars, for example, assume that the strain is constant over the duration of the measurement.
An observation of non-trivial time-dependence could be evidence of new physics.

In addition to the astrophysical motivation, the algorithm provides useful information for detector characterization.
Recent work has shown that geophysical effects such as global Schumann resonances can induce correlated noise in worldwide networks of gravitational-wave detectors in a way that can bias cross-correlation searches~\cite{schumann,wsubtract}.
Some geophysical phenomena (such as electromagnetic activity induced by solar flares) exhibit temporal fluctuations on very long timescales.
Understanding the time dependence of excess cross correlation can help to rule out (or lend credence to) different explanations of an apparent signal in searches for persistent gravitational waves in both targeted radiometer searches~\cite{sph_results} and searches for stochastic backgrounds~\cite{stoch-S5}.

In order to see how very long signals might escape detection without a dedicated search, we note simple scaling behavior.
Radiometric signal-to-noise ratio (SNR, defined below; or see Eq.~3.11 of~\cite{stamp}) scales like 
\begin{equation}\label{eq:snr0}
\text{SNR} \propto \zeta \, t_\text{obs}^{1/2} ,
\end{equation}
where $t_\text{obs}^{1/2}$ is the observation duration and $\zeta$ is the fraction of the observation during which the signal is present.
To be concrete, let us consider an $\approx$$\unit[11]{hr}$-long signal with  $\text{SNR}_\text{tot}=40$.
While this signal is exceptionally loud, it can be significantly diluted if we search for it assuming a persistent signal model.
For example, if we integrate over the entirety of a $\unit[90]{day}$ science run, the signal-to-noise ratio will be diluted by a factor
$$
\frac{(1)\sqrt{\unit[11]{hr}}}{(\unit[11]{hr}/\unit[90]{days})\sqrt{\unit[90]{days}}} = 
\sqrt{\frac{\unit[90]{days}}{{\unit[11]{hr}}}} = 14 ,
$$
so that $\text{SNR}_\text{tot}$ is reduced from $40$ to $\approx$$3$, which is not statistically significant given a few thousand independent frequency bins.
It follows that a very long transient signal, loud enough to affect a radiometer or stochastic search, should produce an easily identifiable signal using an algorithm dedicated to detection of very long transient phenomena.

If we attempt to resolve the signal using $\unit[1]{s}$ segments, the signal-to-noise ratio is diluted by a factor of
$$
\frac{(1)\sqrt{\unit[11]{hr}}}{(1)\sqrt{\unit[1]{s}}} \approx 200 ,
$$
so that $\text{SNR}\approx0.20$ in each segment.
This back-of-the-envelope argument illustrates how even exceptionally loud very long signals can escape detection without a dedicated search.

\section{Radiometry}\label{radiometry}
The basic idea of radiometry is to cross-correlate data from two or more detectors.
By integrating over as much data as possible, it is possible to identify weak signals buried in noise.
Drawing on previous results~\cite{stamp,stochtrack,stochsky}, we take as our starting point spectrograms of cross-correlated data.
Details are provided in Appendix~\ref{details}.
For the sake of simplicity, we focus here on a two-detector network, but note that the algorithm generalizes straightforwardly to arbitrarily many detectors~\cite{stamp}.
The cross-correlated data can be represented as a complex-valued signal-to-noise ratio~\cite{stochsky} (see Eq.~\ref{eq:complex} in Appendix~\ref{details}):
\begin{equation}\label{eq:complex0}
  \mathfrak{p}(t;f) = \kappa\,
  \tilde{s}_1^*(t;f) \tilde{s}_2(t;f) \Big/ \sqrt{P'_1(t;f) P'_2(t;f)} .
\end{equation}
Here, $\tilde{s}_I(t;f)$ is the discrete Fourier transform of the strain series from detector $I$, $P'(t;f)$ is the auto-power spectrum calculated using neighboring data segments, and $\kappa$ is a normalization factor; (see Eq.~\ref{eq:kappa} in the Appendix).
Each segment has duration $\delta t$.
The choice of $\delta t$, which we examine in greater detail below, is determined both by astrophysical considerations and computational limitations.
The argument $t$ is the mid time of the data segment and $f$ is frequency.
Note that $\mathfrak{p}(t;f)$ does not depend on the direction of the source $\hat\Omega$ because the direction of the source is encoded as a complex phase.

\begin{figure*}[hbtp!]
  \subfigure[]{\psfig{file=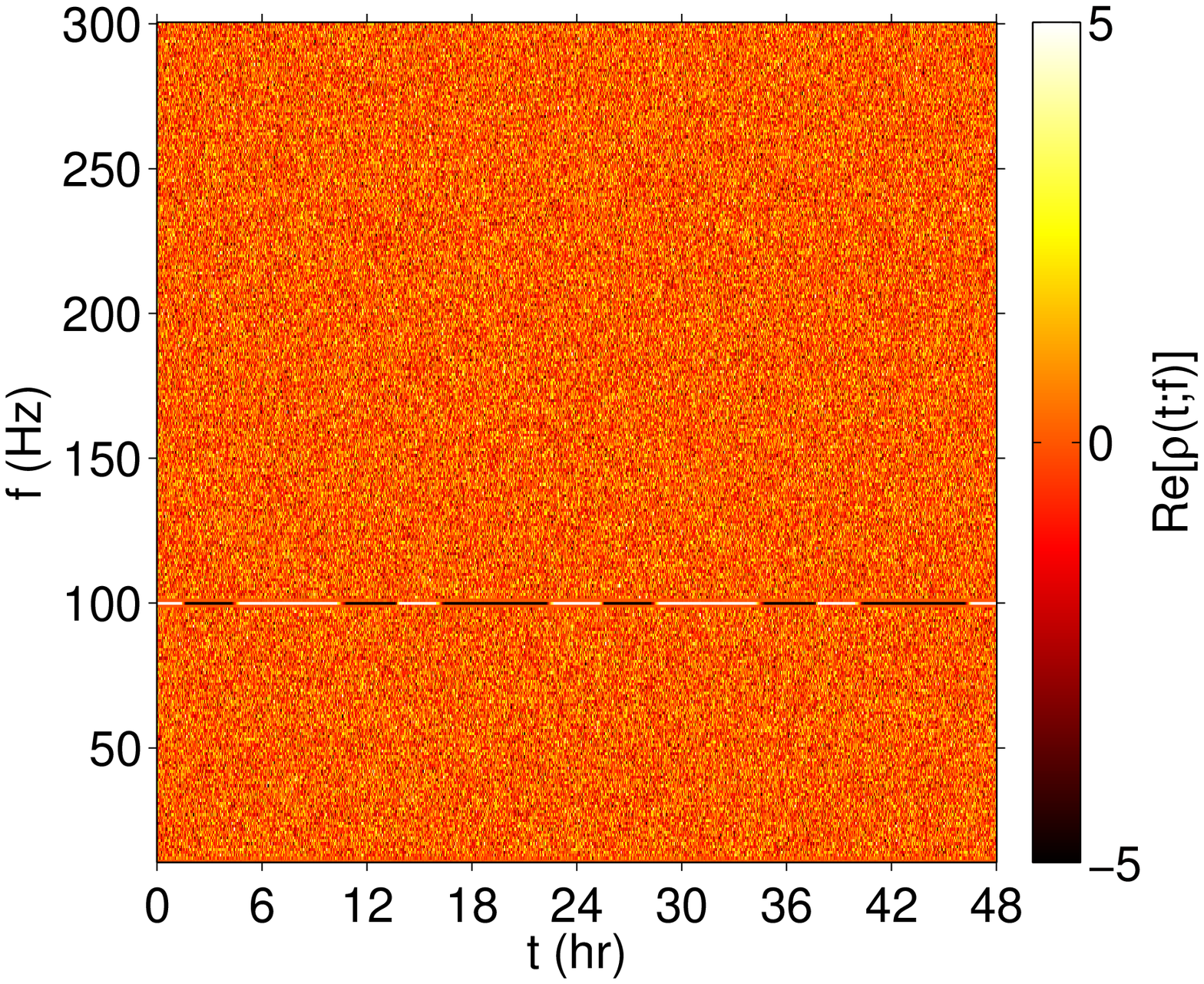, height=2.8in}}
  \subfigure[]{\psfig{file=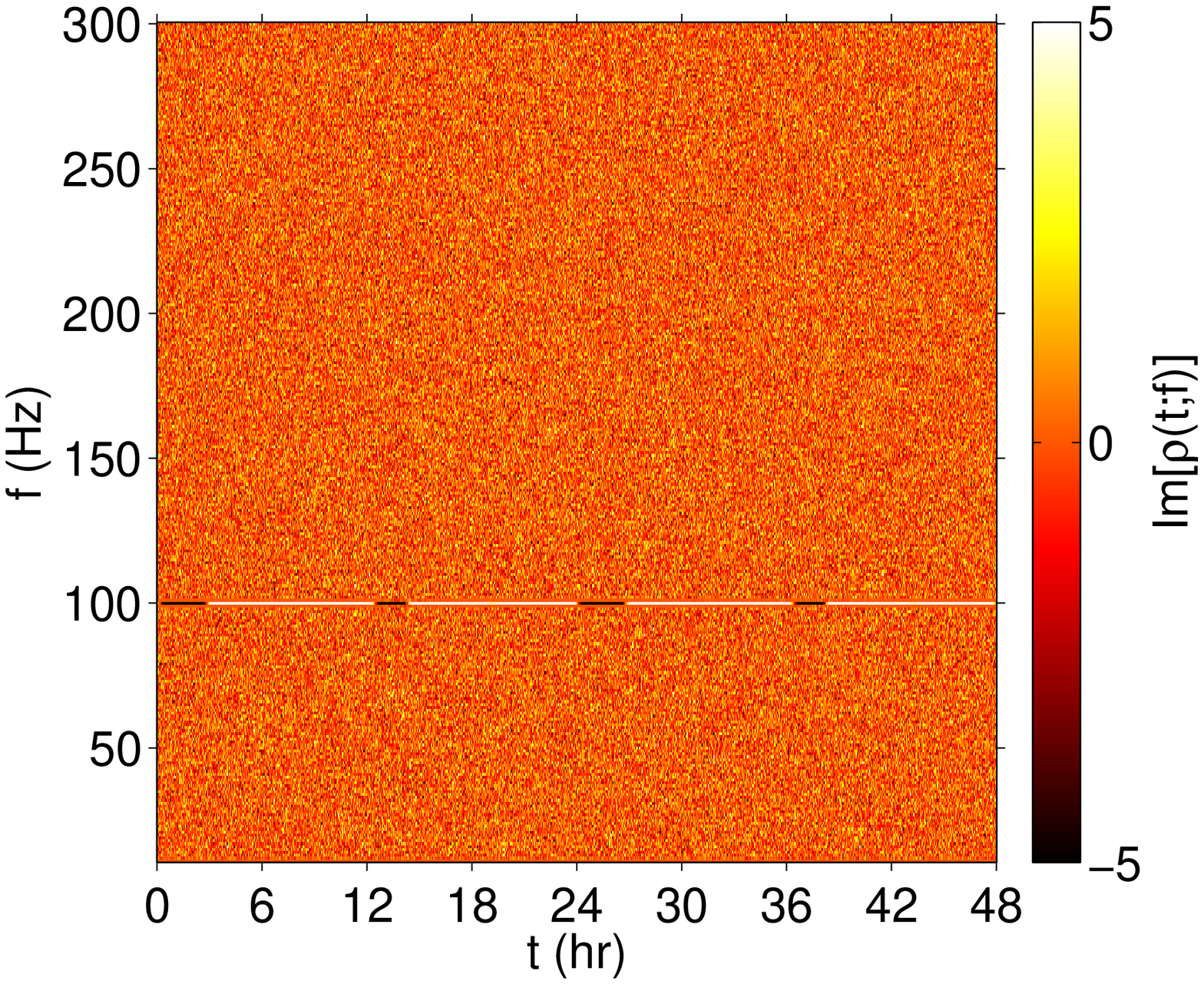, height=2.8in}}
  \subfigure[]{\psfig{file=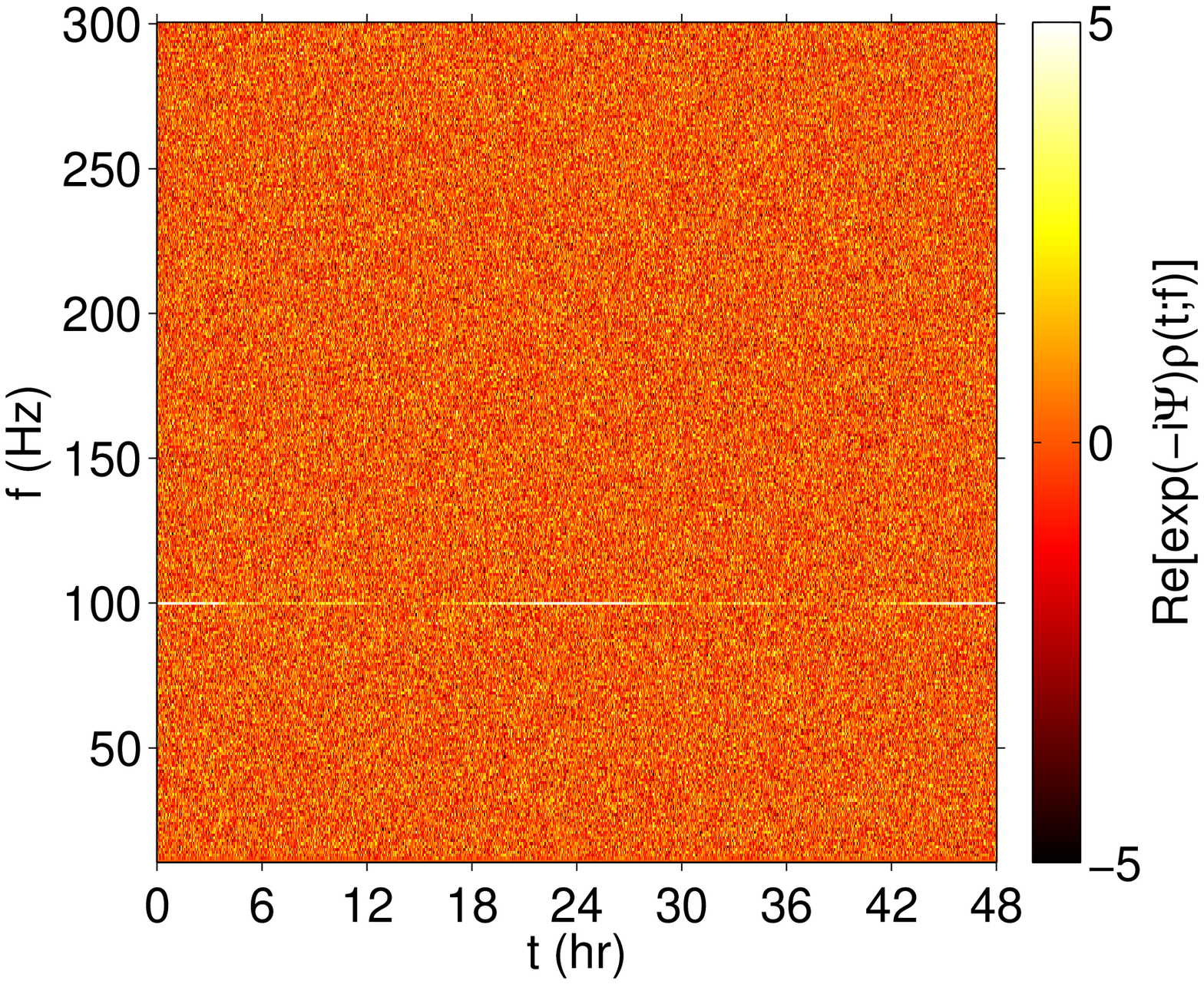, height=2.8in}}
  \subfigure[]{\psfig{file=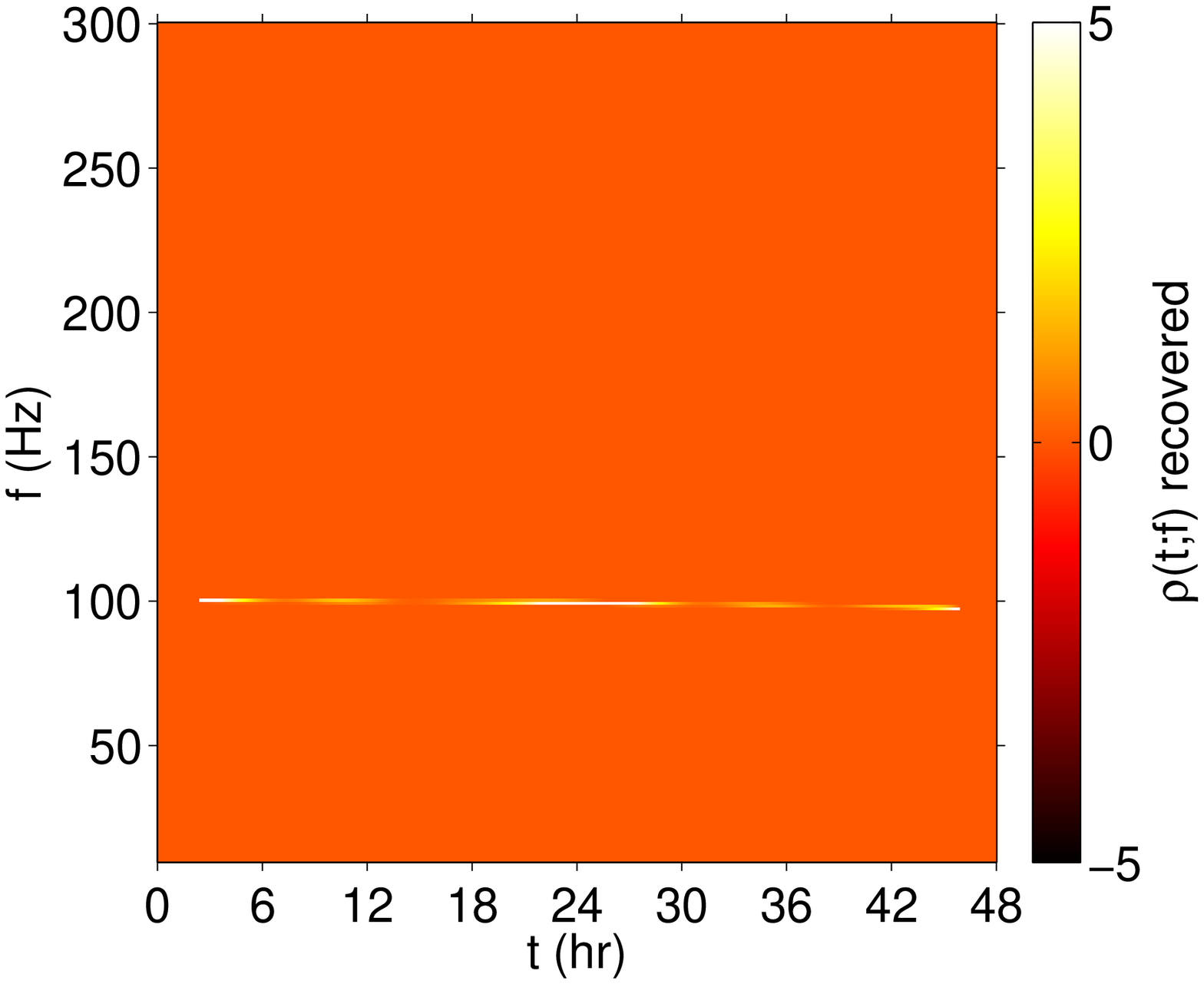, height=2.8in}}
  \caption{
    The daily modulation of $\mathfrak{p}$.
    Each panel shows a $(\delta t, \Delta f)=(\unit[158]{s},\unit[1]{Hz})$ spectrogram consisting of two consecutive days of data.
    The data consist of Advanced LIGO~\cite{aligo} Monte Carlo noise plus a simulated signal at $f=\unit[100]{Hz}$.
    The top-left panel (a) shows $\text{Re}(\mathfrak{p})$ while the top-right panel (b) shows $\text{Im}(\mathfrak{p})$.
    In the top panels, the injected signal, for illustrative purposes, is very loud:  $h_0=1\times10^{-22}$.
    (Since the signals are very loud, the color bar is saturated.)
    At $t=0$, the phase angle is in the lower-right quadrant of the $\mathfrak{p}$ complex plane where $\text{Re}(\mathfrak{p})>0$ and $\text{Im}(\mathfrak{p})<0$.
    It rotates clockwise so that $\text{Re}(\mathfrak{p})$ crosses zero and becomes negative before $\text{Im}(\mathfrak{p})$ crosses zero and becomes positive.
    The bottom-left panel (c) shows $\text{Re}\left[\exp\left(-i\Psi_i\right) \mathfrak{p}\right]$ for a much weaker signal: $h_0=4\times10^{-24}$.
    The phase factor is applied to make the signal entirely real; see Eq.~\ref{eq:Y}.
    The bottom-right panel (d) shows the recovered signal (the most significant cluster identified in the bottom-left panel) recovered with $2\times10^6$ seedless clustering templates (see Section~\ref{demonstration} for details).
    The source is located at $(\text{ra},\text{dec})=(\unit[18.5]{hr},+39^\circ)$ and it is recovered $7^\circ$ degrees away at $(\unit[18.9]{hr},+35^\circ)$.
    The recovery algorithm does not assume that the signal is monochromatic.
    \label{fig:days}
  }
\end{figure*}

If the data consist of well-behaved noise, both the real and imaginary parts of $\mathfrak{p}$ are characterized by distributions with mean$=$0 and variance$\approx$1.
Gravitational-wave signals induce excess $\mathfrak{p}$ with a complex phase angle $\Psi$ determined by the direction of the source $\hat\Omega$ and its frequency $f_i$:
\begin{equation}\label{eq:Psi}
  \Psi_i \equiv -2\pi f_i \hat\Omega \cdot \Delta \vec{x}/c
\end{equation}
Here, $\Delta\vec{x}$ is the difference in detector position and $c$ is the speed of light.
For the LIGO detectors, $|\Delta\vec{x}|/c\approx\unit[10]{ms}$.
Thus, a signal can, e.g., induce real negative or positive imaginary values of $\mathfrak{p}$.
Note that $\Psi$ is time-dependent due to the rotation of the Earth.
This behavior is illustrated in Figs.~\ref{fig:days}a and~\ref{fig:days}b, which show spectrograms of the real and imaginary parts of $\mathfrak{p}$ for data consisting of Advanced LIGO~\cite{aligo}, Gaussian, Monte Carlo noise at design sensitivity plus a persistent monochromatic signal with frequency $f=\unit[100]{Hz}$ and strain amplitude $h_0=1\times10^{-22}$, located at $(\text{ra},\text{dec})=(\unit[18.5]{hr},+39^\circ)$.

These spectrograms use $(\delta t, \Delta f)=(\unit[158]{s},\unit[1]{Hz})$ resolution and span two days of data.
We utilize a coarse-graining procedure described below.
The repeating pattern of light and dark at $\unit[100]{Hz}$ repeats with a period of one sidereal day.
This pattern, which varies with declination and frequency, shows how the phase factor $\Psi_i$ (Eq.~\ref{eq:Psi}) varies with the rotation of the Earth.
It is interesting to note that these horizontal stripes are a distinct but analogous effect to the vertical stripes (also arising from radiometer phase mismatch) described in~\cite{stochsky}.

If we multiply $\mathfrak{p}$ by (an array of) the phase factor $\exp{(-i\Psi_i)}$ in order to ``point'' the radiometer map in the source direction (and reduce the injected strain amplitude to $h_0=4\times10^{-24}$) we obtain Fig.~\ref{fig:days}c.
With the appropriate phase factor applied, the cross-power due to signal is real and positive.
However, there is still a time-dependent modulation with a period of one sidereal day, owing to the time dependence of the detectors' antenna response; see Eq.~\ref{eq:epsilon}.
Fig.~\ref{fig:days}d shows the signal recovered using a seedless clustering algorithm~\cite{stochsky} to look for tracks of bright pixels.
This recovery is described in greater detail in Section~\ref{demonstration}.

In order to carry out a search for a transient signal associated with some direction $\hat\Omega$, we construct a detection statistic (see Eq.~\ref{eq:vlong_as} in Appendix~\ref{details}):
\begin{equation}\label{eq:vlong_as0}
  \text{SNR}_\text{tot}(\hat\Omega) = \frac{\text{Re}\left[
    \sum_{i\in\Gamma}
    \exp\left(-i \Psi_i\right)
    \mathfrak{p}_i \, \epsilon_{12}^i
    \right]
  }{
    \left(\sum_{i\in\Gamma}\Big(\epsilon_{12}^i(\hat\Omega)\Big)^2\right)^{1/2}
  } .
\end{equation}
The $\epsilon_{12}^i(\hat\Omega)$ factor describes the efficiency of the detector pair~\cite{stamp} (see Eq.~\ref{eq:epsilon} in Appendix~\ref{details}).
The sum over $i$ runs over some set of $(t;f)$ pixels $\Gamma$, which is determined by details of the search.
A search for persistent narrowband signals, for example, might sum over time at a fixed frequency.
In searches for transient signals, the set of $(t;f)$ pixels may be described, e.g., by a B\'ezier curve~\cite{stochtrack,stochsky,stochtrack_cbc}.
In the analysis that follows, we employ clustering with parametrized quadratic B\'ezier curves, each with a different randomly chosen direction.
This family of curves includes straight lines as a subset.
For fixed $\hat\Omega$, $\epsilon_{12}^i(\hat\Omega)$ varies as the Earth rotates.
While this effect is usually not important on timescales of $\lesssim$$\unit[1000]{s}$, it plays a significant role for very long signals.
Finally, note that Eq.~\ref{eq:vlong_as0} assumes (for the sake of compact notation) that the noise is approximately stationary and that the signal does not vary significantly with frequency; see Eq.~\ref{eq:rho2}.
These assumptions can be straightforwardly relaxed using Eq.~\ref{eq:snr}.

\section{Compression}\label{compression}
\subsection{Coarse-graining}
In order to facilitate the search for very long-lived signals, it is desirable to compress the cross-correlation data used in long transient searches, e.g.,~\cite{lgrb}.
Past analyses, targeting signals with duration $\approx$$10$--$\unit[1000]{s}$, utilized ($50\%$ overlapping) $\unit[1]{s}$-long segments.
While this does not pose a significant computational challenge for the analysis of a $\unit[1500]{s}$-long window ($\unit[3000]{segments}$), it is not practical for the analysis of $\unit[10]{hr}=\unit[36000]{s}$ signals ($\unit[72000]{segments}$).
First, it is currently impractical to load the data from so many segments into RAM.
Second, even if we could get around the memory problem, it would take an inordinate amount of computation time to treat so many segments separately.

One solution is to average data from neighboring frequency bins in order to compress the data into a format that preserves the information needed to search for very long signals while averaging away details about short timescales.
This compression, known as ``coarse-graining,'' enforces the assumption that the signal varies slowly enough that the change in frequency as a function of time is small compared to the bandwidth of each spectrogram pixel.
Previous radiometer analyses~\cite{radiometer,sph_results} have already employed a form of coarse-graining as part of a standard data conditioning procedure, but in these cases, the end data product is a single spectrum with no time-dependence (as opposed to a spectrogram), and so the coarse-graining was somewhat incidental.

We follow the coarse-graining procedure first outlined in~\cite{stoch_s1} (see Eq.~5.4), which we reproduce here in slightly different notation.
First, we define cross-power $C(t;f)\equiv(2/{\cal N})\tilde{s}_1^*(t;f) \tilde{s}_2(t;f)$ where ${\cal N}$ is a Fourier normalization constant; see Eq.~\ref{eq:kappa} in Appendix~\ref{details}.
This allows us to rewrite Eq.~\ref{eq:complex0} as~\footnote{In our calculation of discrete Fourier transforms, we employ a Tukey window with a $\unit[0.5]{s}$ rise and fall time.}:
\begin{equation}
  \mathfrak{p}(t;f) = \sqrt{2}\, C(t;f) \Big/ \sqrt{P'_1(t;f) P'_2(t;f)} .
\end{equation}
The coarse-grained cross-power is given by:
\begin{equation}\label{eq:CG}
    C_\text{CG}(t;f_\text{CG}) \equiv \frac{\delta f}{\Delta f} \sum_{f=f_\text{CG}-\Delta f/2}^{f=f_\text{CG}+\Delta f/2} w(f) \, C(t;f|\hat\Omega) .
\end{equation}
Here, $\Delta f$ is the width of the new coarse-grained bins, $\delta f$ is the original fine-grained bin width, and $f_\text{CG}$ are the new coarse-grained frequency bins.
The variable $w(f)$ is a weight factor, which is unity when $f\neq f_\text{CG}\pm\Delta f/2$.
When $f=f_\text{CG}\pm\Delta f/2$, then $w(f)$ is some number on $[0,1]$, which takes into account the fact that the edges of a coarse-grained bin may fall in between the edges of two fine grained bins.
The precise value is given by the fraction of the fine-grained bin that overlaps with the coarse-grained bin.
Thus, e.g., a $\Delta f=\unit[0.25]{Hz}$ coarse-grained bin spanning $10.0$--$\unit[10.25]{Hz}$ would include a $w=0.5$ overlap with a $\delta f=\unit[1/60]{Hz}$ fine-grained bin spanning $9.98$--$\unit[10.017]{Hz}$.

Eq.~\ref{eq:CG} can be understood as an application of Parseval's theorem.
The analogous coarse-grained auto-powers, $P'_{1\,\text{CG}}(t;f)$, $P'_{2\,\text{CG}}(t;f)$ are calculated using Welch's method~\cite{welch}.
We thereby obtained the coarse-grained estimator:
\begin{equation}
  \mathfrak{p}_\text{CG}(t;f) = \sqrt{2}\, C_\text{CG}(t;f) \Big/ \sqrt{P'_{1\,CG}(t;f) P'_{2\,CG}(t;f)} .
\end{equation}

In addition to Eq.~\ref{eq:CG}, we considered an alternative coarse-graining procedure performed in the time domain, which we refer to as ``hybrid compression.''
Instead of combining many narrow frequency bins associated with long segments, we combined many short segments associated with wide frequency bins.
While the two procedures produced similar results, we find that the frequency-domain implementation reduced spectral leakage by up to tens of percent in power compared to the time-domain implementation, thereby allowing for better signal reconstruction.
That said, there may be applications in which it it useful to carry out hybrid compression: first in the frequency domain and then in the time domain.
We return to this idea at the end of the next subsection.

\subsection{Limitations}
The rotation of the Earth places a limit on the extent of possible compression~\footnote{
An independent, but weaker constraint arises by taking into account Doppler modulation.
In the case of a half-year-long signal (for which Doppler modulation is maximal), the modulation depth is $\approx 10^{-4} \, f$.
Thus, a radiometer search for months-long signals at $f\approx\unit[1000]{Hz}$ should employ a frequency resolution $\Delta f \gtrsim\unit[0.1]{Hz}$.
}.
If the segment duration $\delta t$ is too long, the Earth will rotate significantly during the duration of the segment so that it becomes impossible to ``point'' the radiometer in different directions by weighting each spectrogram pixel with the appropriate phase factor; see~\cite{stochsky}.
In practice, this translates to a loss of signal-to-noise ratio.

Following~\cite{stochsky}, the fraction of signal-to-noise ratio {\em not} lost due to phase offset $R$ can be written as
\begin{equation}\label{eq:R}
  R = \cos(2 \pi f \Delta\vec{x} \cdot \Delta\hat\Omega /c ) ,
\end{equation}
where $f$ is gravitational-wave frequency, $\Delta\vec{x}$ describes the displacement between the two detectors, $\Delta\hat\Omega$ describes the angular error due to the rotation of the Earth, and $c$ is the speed of light.

The angular error depends on the segment duration relative to a sidereal day.
The local coordinates for directions corresponding to the poles ($\text{dec}=\pm90^\circ$) do not change at all as the Earth rotates.
The change is greatest for directions in the equatorial plane $\text{dec}=0^\circ$, for which
\begin{equation}
  \left|\Delta\hat\Omega\right| = 2\pi \frac{\delta t / 2}{\unit[24]{hr}} .
\end{equation}
Here, $\delta t$ is the segment duration.

If we conservatively assume that the source is located at $\text{dec}=0^\circ$, we can rewrite Eq.~\ref{eq:R} as
\begin{equation}
    R = \cos\left[
      2 \pi f |\Delta\vec{x}/c| 
      (\pi \delta t / \unit[24]{hr})
      \cos(\psi) 
      \right] ,
\end{equation}
where $\psi$ is the angle between $\Delta\vec{x}$ and $\Delta\hat\Omega$.
The loss of signal-to-noise ratio is (conservatively) maximized when $\cos(\psi)=1$.
Thus, we obtain the following expression for $\delta t$:
\begin{equation}
  \delta t =
  \frac{\cos^{-1}(R)}{
    2 \pi f \, t_0
    (\pi / \unit[24]{hr})
  } ,
\end{equation}
where $t_0\equiv |\Delta\vec{x}/c|$ is the travel time between the two detectors ($\approx$$\unit[10]{ms}$ for the Hanford-Livingston pair).
If we demand that $R>90\%$, we obtain
\begin{equation}\label{eq:limit}
  \delta t \lesssim \unit[197]{s} \left(\frac{\unit[1000]{Hz}}{f} \right)
\end{equation}
for the Hanford-Livingston detector pair.
Employing a more conservative requirement of $R>95\%$, the prefactor becomes $\unit[140]{s}$.

\subsection{Cost}
Now, we compare the computational cost of an analysis using uncompressed data, consisting, e.g., of 50\%-overlapping $\unit[1]{Hz}\times\unit[1]{s}$ spectrogram pixels, to the cost of an analysis using compressed data consisting of non-overlapping $\unit[1]{Hz}\times\unit[158]{s}$ pixels.
After compression, a $\unit[10]{hr}$-long stretch of data is described by only $228$ (non-overlapping) $\unit[158]{s}$ coarse-grained segments~\footnote{Overlapping segments are often used in long-transient searches to recover signal power lost to Hann windowing, which, in turn, is carried out to reduce spectral leakage.  
Here, we have replaced Hann windows with a rapidly rising Tukey window.
This allows us to minimize spectral leakage without a significant reduction in signal power, thereby facilitating the use of non-overlapping segments.
}.
This makes it straightforward to search for $\sim$day-long signals.
One can analyze signals with durations in excess of one week e.g., by splitting the analysis band into overlapping sub-bands.

Further, having reduced the size of the dataset by a factor of $2(\unit[158]{s})/\unit[1]{s}\approx$$316$, an all-sky search using seedless clustering will run $\approx$$316$ times faster (than an analysis using uncompressed, $50\%$-overlapping, $\unit[1]{Hz}\times\unit[1]{s}$ spectrograms).
Seedless clustering uses phaseless templates to approximate narrowband gravitational-wave signals~\cite{stochtrack}.
The sensitivity of the search, as well as its cost, is influenced by the number of templates~\cite{stochsky}.

The cost of a radiometer search is usually dominated by background estimation.
The standard procedure relies on time slides, in which one data stream is repeatedly shifted in time with respect to the other in order to obtain many realizations of detector noise.
Actual gravitational-wave signals do not appear in time-shifted data since the coherence is destroyed.
In order to detect signals with a $3\sigma$ false alarm probability of $\approx$0.3\%, it is necessary to analyze the $\approx$370 realizations of time-shifted data.

Scaling estimates from~\cite{stochsky}, and assuming $\approx$$\unit[3]{day}$-long spectrograms, we estimate the cost of background estimation for a realistic search to the $3\sigma$ level. 
We assume the search is  carried out with $T_{150}=2\times10^6$ templates per $\unit[150]{Hz}$ of bandwidth (a typical number of templates~\cite{stochsky}).
We find that it can be carried out using $3.6$ continuously-running eight-core Intel Xeon E5-4650 CPUs or Kepler GK104 GPUs~\cite{stochsky}.
This translates to $29$ continuously-running CPU cores.
The computational cost for background estimation in a realistic search scales as follows:
\begin{equation}\label{eq:cost}
  \begin{split}
    t_c \approx & \unit[10]{days}
    \left(\frac{T_{150}}{2\times10^6}\right)
    \left(\frac{t_\text{obs}}{\unit[1]{yr}}\right)
    \left(\frac{B}{\unit[1200]{Hz}}\right) \\
    &
    \left(\frac{n_\text{ts}}{370}\right)
    \left(\frac{128}{n_\text{GPU}}\right)
    \left((2)158\Big/\frac{\Delta f}{\delta f}\right)
  .
  \end{split}
\end{equation}
Here, $t_c$ is the computation time, $t_\text{obs}$ is the duration of the observing run, $B$ is the width of the analysis band, and $n_\text{ts}$ is the number of time slides (in this case, corresponding to three-sigma confidence).
The variable $n_\text{GPU}$ is the number of GPUs (or, equivalently, 8-core CPUs).
The ratio of $(\Delta f/\delta f)$ describes the compression achieved with coarse-graining.
For the examples shown here, $(\Delta f/\delta f)=158$.
The factor of two is included if we replace overlapping windows with non-overlapping windows.

\subsection{Hybrid compression}
At the end of the previous subsection, we noted that there may be applications for which it is helpful to carry out hybrid compression: first in the frequency domain and then in the time domain.
For example, if we are interested a particular direction in the sky, one can combine many frequency-domain coarse-grained segments into even longer segments via Eq.~\ref{eq:vlong_as0}.
Since, by assumption, each of the frequency-domain coarse-grained segments in the sum already takes into account the phase offset between the two detectors, the constraint from Eq.~\ref{eq:limit} does not apply, and so there is no limit to the number of segments that can be combined.
In this way, it should be possible to represent an arbitrarily long span of data with a fixed number of spectrogram pixels, facilitating a search for a signal of any duration.
It is important to note, however, that this hybrid technique can only be applied by assuming a specific source location, and so it does not lend itself to an efficient all-sky search.

\subsection{Trade-offs}
By coarse-graining spectrographic data with Eq.~\ref{eq:CG}, we are trading reduced frequency resolution for reduced computational burden.
As a consequence, the sensitivity of the search is reduced in comparison to an idealized lossless search using arbitrarily large computing resources.
To see why, consider a narrowband signal passing through just one fine-grained spectrogram pixel (width = $\delta f$) during each segment in time.
When the data are coarse-grained to a width $\Delta f$, the signal in this single fine-grained bin is averaged with data in neighboring frequency bins containing only noise.
In this worst-case scenario, the detection statistic $\text{SNR}_\text{tot}$ is reduced by a factor $\sqrt{\Delta f / \delta F}$---corresponding to a factor of $\approx$$13$ for the $\unit[1]{Hz}\times\unit[158]{s}$ pixels used for our demonstration below.

In practice, the loss of sensitivity can be less if the signal is not confined to a single fine-grained bin at any given time.
Also, fine-grained bins are sometimes contaminated with noise artifacts including non-Gaussian and non-stationary noise, which are, to some extent, ameliorated through coarse-graining.
Such noise artifacts can therefore serve to make the loss of sensitivity less severe in comparison to the idealized case.

Once the data are coarse-grained, the high-resolution information is lost; the process is not invertible.
However, in the event of a candidate detection, one can imagine reanalyzing the original lossless dataset with a targeted search, looking only in a small band and in a specific direction identified by the coarse-grained search.
By severely restricting the parameter space of the search, it is possible to carry out a computationally feasible follow-up study using fine-grained data with reasonable computational resources.

\section{Demonstration}\label{demonstration}
Our goal here is to show that very long signals, lasting hours or longer, can be successfully identified by employing an existing pattern recognition algorithm on compressed data.
For our first demonstration, we recover the signal shown in Fig.~\ref{fig:days}c.
We create a simulated dataset spanning two days of Monte Carlo data for the Advanced LIGO Hanford and LIGO Livingston detectors operating at design sensitivity.
The data consists of Gaussian noise plus a simulated monochromatic signal with frequency $f=\unit[100]{Hz}$ and strain amplitude $h_0=4\times10^{-24}$, located at $(\text{ra},\text{dec})=(\unit[18.5]{hr},+39^\circ)$.
The signal is circularly polarized.

The data are cross-correlated to produce spectrograms with resolution of $(\delta t, \Delta f)=(\unit[158]{s},\unit[1]{Hz})$ with non-overlapping segments.
We consider an observing band spanning $10$--$\unit[300]{Hz}$.
This spectrogram (with a phase factor applied to point the radiometer in the appropriate direction) is shown in Fig.~\ref{fig:days}c.
Once the data are compressed, we carry out a seedless clustering search~\cite{stochtrack,stochsky}.

We make no assumptions about the start time, stop time, frequency, or sky location of the search---subject to the constraint that the signal morphology can be approximated as a cubic B\'ezier curve of at least $\unit[1.8]{hr}$ in duration.
(Following~\cite{stochsky}, we take advantage of the fact that $\mathfrak{p}$ encodes directional information as a complex phase in order to point each seedless template in a different random direction.)
We record $\text{SNR}_\text{tot}$ for $2\times10^6$ randomly-generated templates in order to identify the template that best fits the signal.
The best-fit template (with $\text{SNR}_\text{tot}=56$) is shown in Fig.~\ref{fig:days}d.
The recovery represents a reasonably good match to the injected signal.
(It would be better recovered using a larger number of templates.)
Running on an eight-core CPU, this $\unit[48]{hr}$ stretch of data was analyzed in $\approx$$\unit[550]{s}$.

As a second test, we attempted to recover a signal with a more complicated time-frequency evolution.
The signal turns on $\approx$$\unit[5]{hr}$ after the beginning of data-taking and ends near $t\approx$$\unit[17]{hr}$ during which time the signal evolves according to:
\begin{equation}\label{eq:phi}
  \begin{split}
    \Phi(t) & = \Phi_0 + 
    2\pi\left[f_0\,(t-t_0) + \frac{1}{2}\dot{f}\,(t-t_0)^2 \right] \\
    h_+ & = h_0 \, \cos\left(\Phi(t)\right) \\
    h_\times & = h_0 \, \sin\left(\Phi(t)\right) .
  \end{split}
\end{equation}
We choose $f_0=\unit[100]{Hz}$, $\dot{f}=\unit[20]{Hz\,day^{-1}}$, and $h_0=4\times10^{-24}$.
The signal is added to simulated noise and processed using the same procedure used for the first test injection.

\begin{figure*}[hbtp!]
  \subfigure[]{\psfig{file=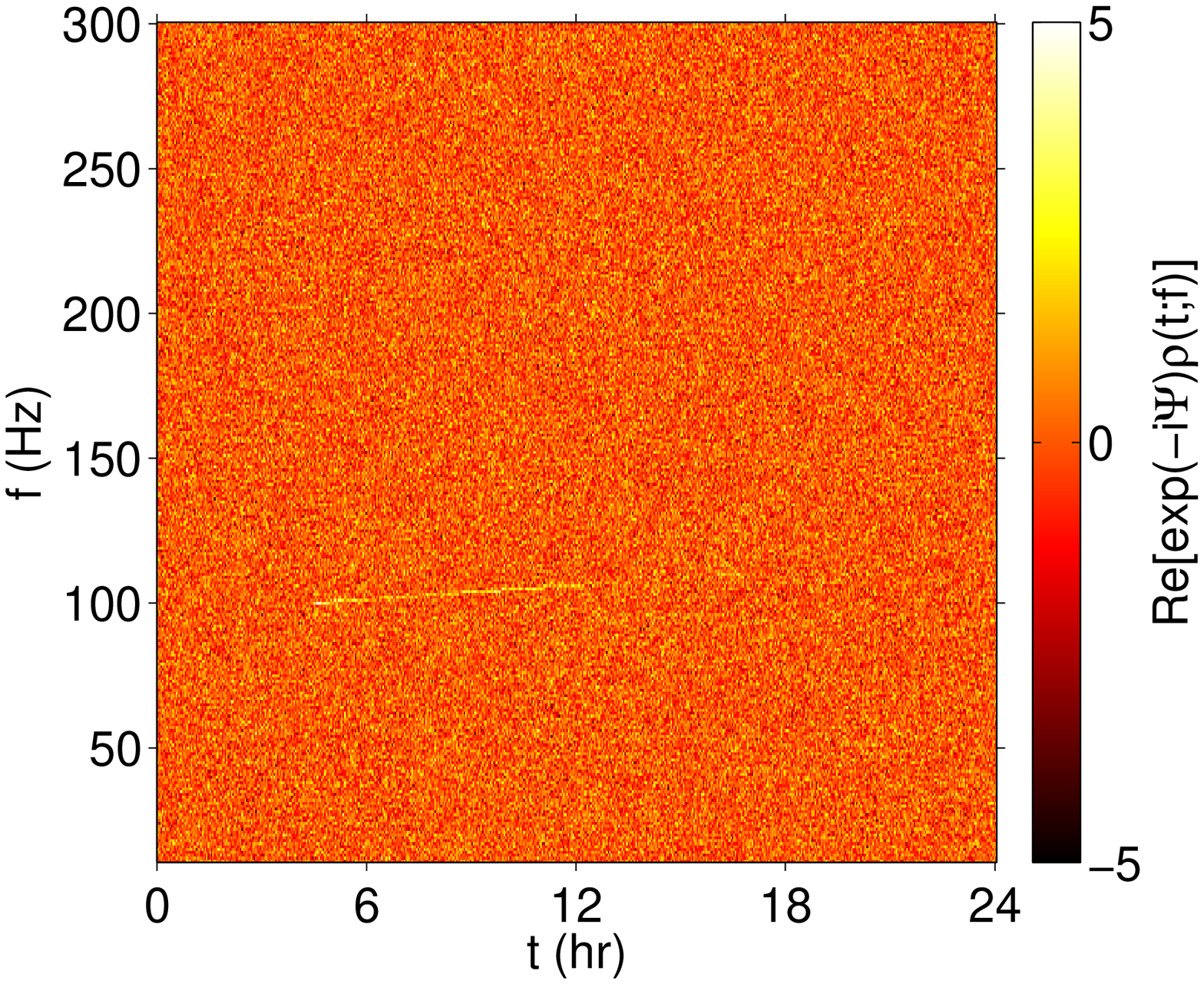, height=2.8in}}
  \subfigure[]{\psfig{file=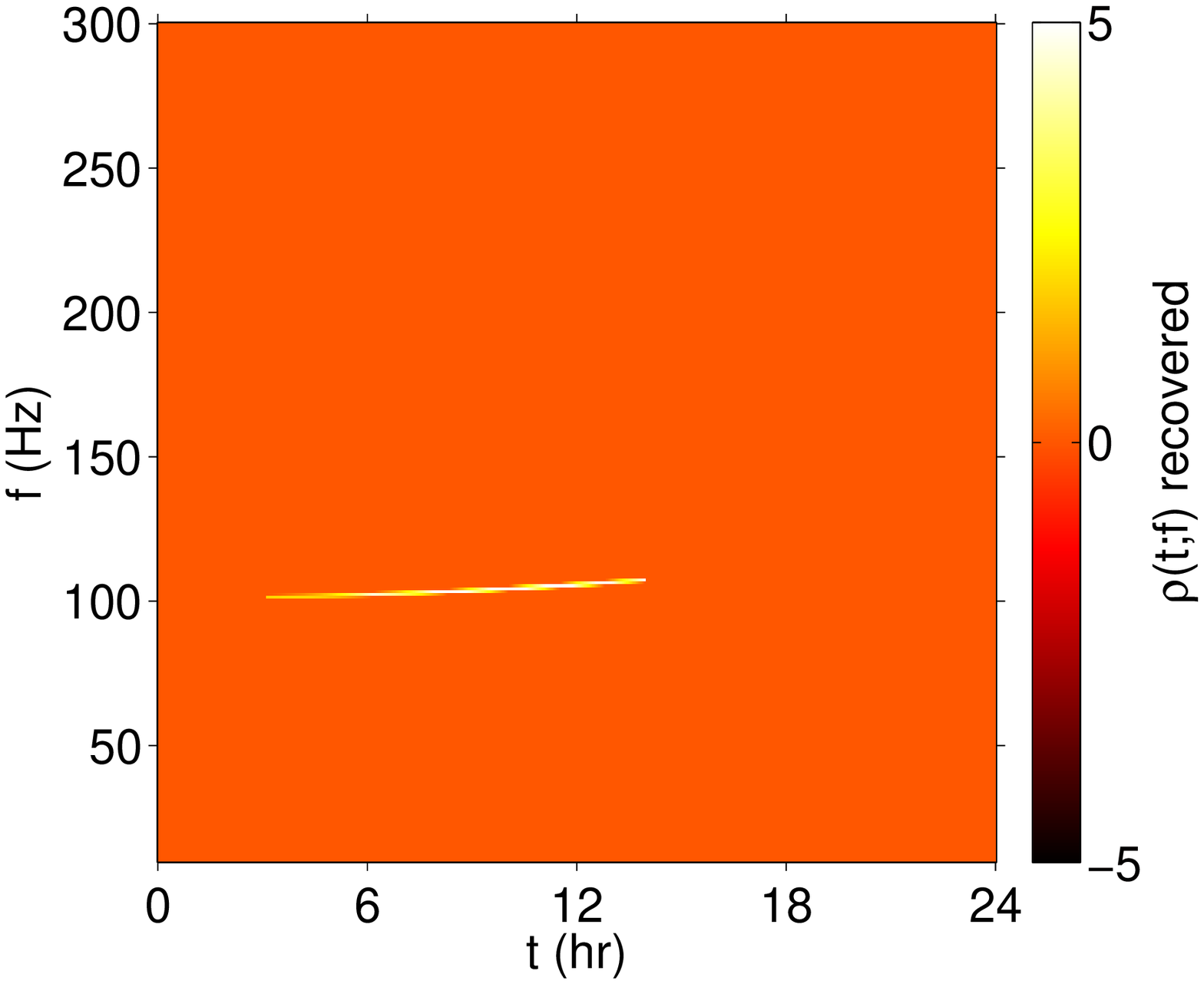, height=2.8in}}
  \subfigure[]{\psfig{file=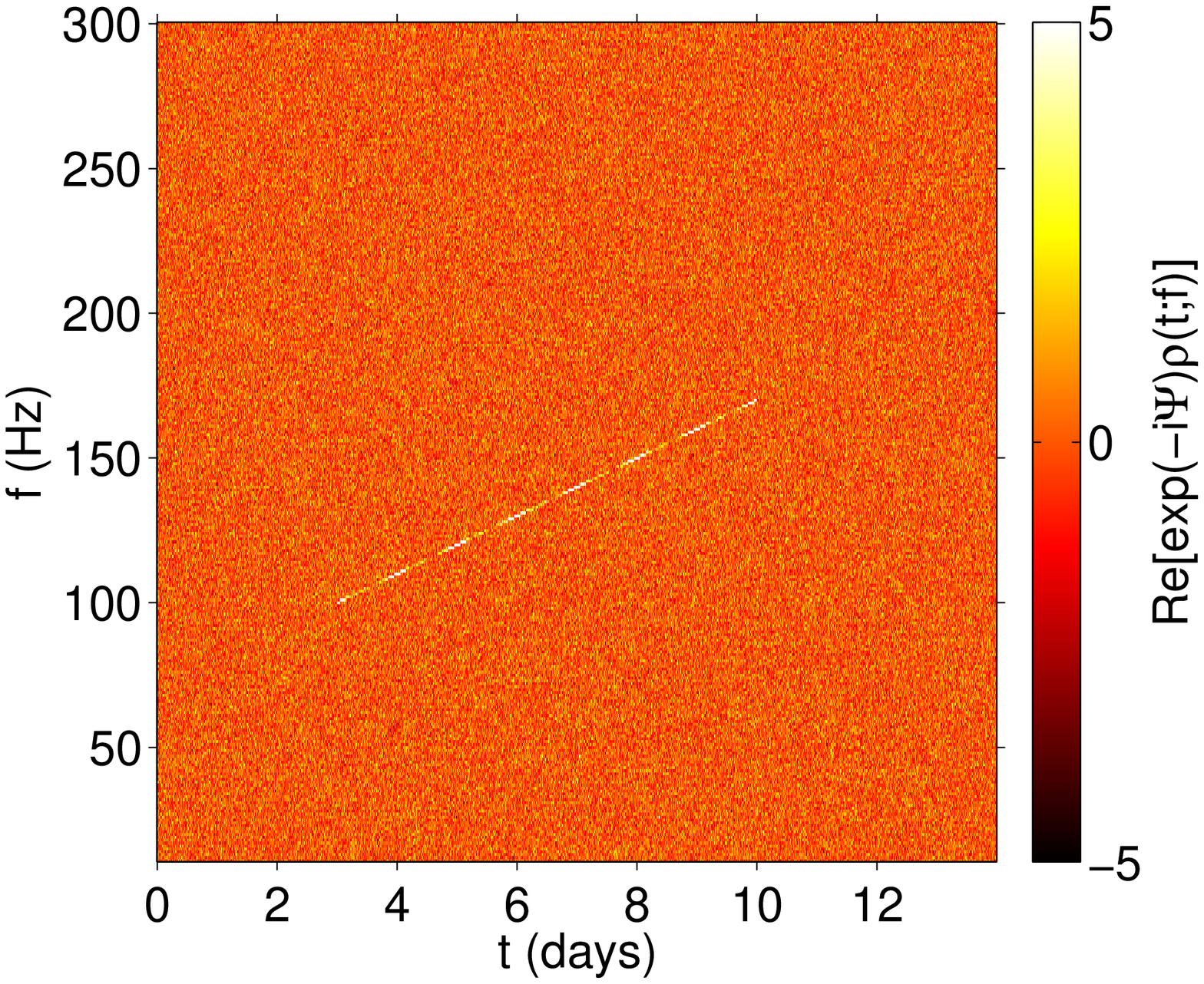, height=2.8in}}
  \subfigure[]{\psfig{file=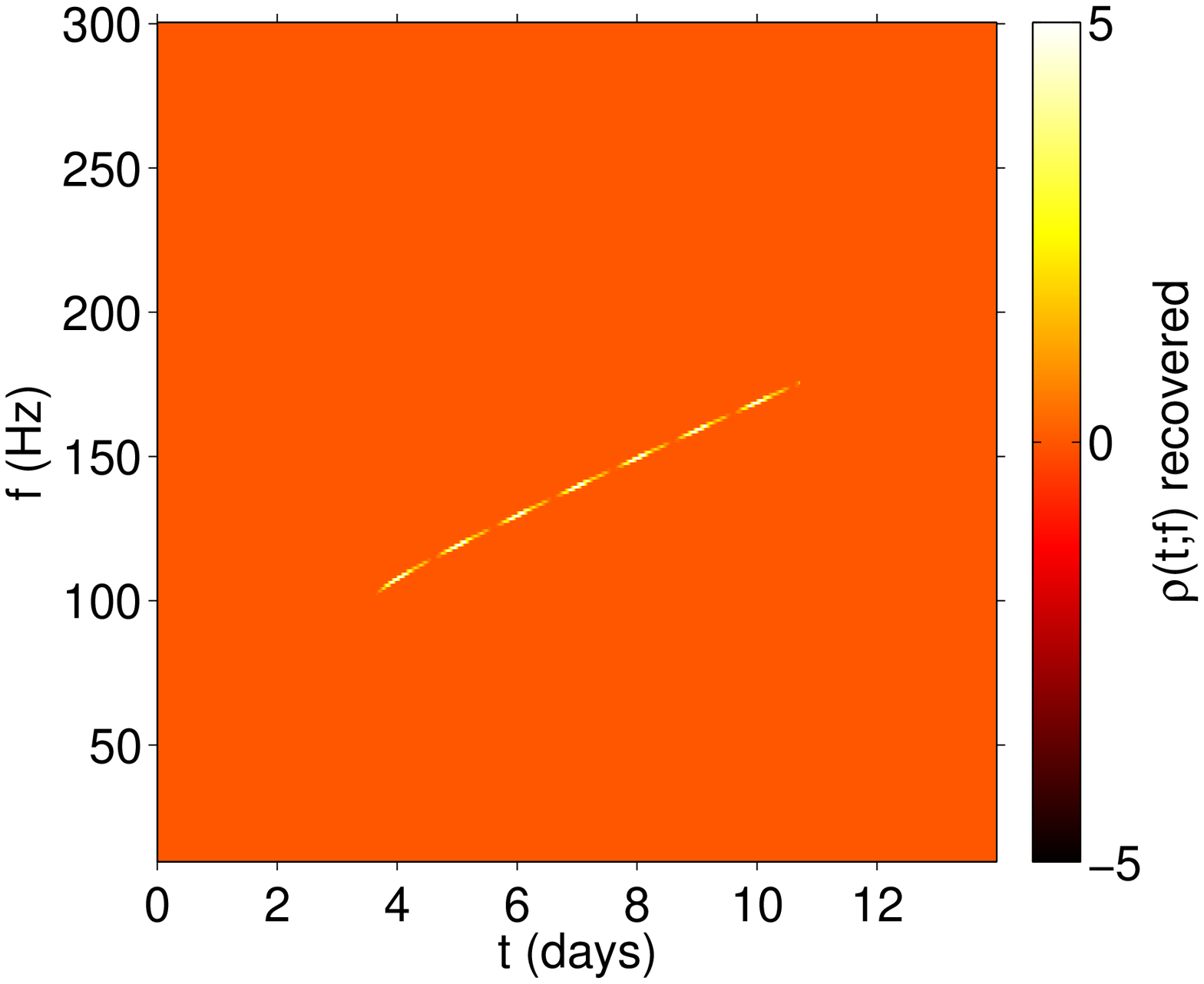, height=2.8in}}
  \caption{
    Recovery of a very long transient signal with seedless clustering.
    Top-left (a): a $(\delta t, \Delta f)=(\unit[158]{s},\unit[1]{Hz})$ spectrogram of $\rho(t;f|\hat\Omega)$ (Eq.~\ref{eq:rho}) consisting of one day of data.
    The data consist of Advanced LIGO Monte Carlo noise plus a simulated signal.
    The signal begins at $f=\unit[100]{Hz}$ and increases with a spin-up given by $\dot{f}=\unit[20]{Hz\,day^{-1}}$.
    The amplitude is $h_0=4\times10^{-24}$.
    The signal persists between $t\approx$$5$--$\unit[17]{hr}$.
    (It grows very faint after $t\approx$$\unit[12]{hr}$ due the unfavorable orientation of the Earth during this period.)
    Top-right (b): the recovered signal recovered with $2\times10^6$ cubic B\'ezier templates; $\text{SNR}_\text{tot}=11$.
    Bottom-left (c): a $(\delta t, \Delta f)=(\unit[158]{s},\unit[1]{Hz})$ spectrogram of $\rho(t;f|\hat\Omega)$ consisting of two weeks of data.
    The data consist of Advanced LIGO Monte Carlo noise plus a simulated signal.
    The signal begins at $f=\unit[100]{Hz}$ and increases with a spin-up given by $\dot{f}=\unit[10]{Hz\,day^{-1}}$.
    The amplitude is $h_0=4\times10^{-24}$.
    The signal persists between $t\approx$$3$--$\unit[10]{days}$.
    Bottom-right (d):  the recovered signal the most significant cluster identified in the bottom-left panel) recovered with $2\times10^6$ cubic B\'ezier templates; $\text{SNR}_\text{tot}=67$.
  }
  \label{fig:inj2}
\end{figure*}

In Fig.~\ref{fig:inj2}a, we show a spectrogram of $\rho(t;f)$.
The signal can be seen as a faint track beginning at $\unit[5]{hr}$ and appearing to end at $\unit[12]{hr}$.
(The last several hours of the signal are difficult to see by eye since the alignment of the detectors makes it weak during this span.)
The recovered signal is shown in Fig.~\ref{fig:inj2}b.
Once again, we use $2\times10^6$ cubic B\'ezier templates.
As before, the source is located at $(\text{ra},\text{dec})=(\unit[18.5]{hr},+39^\circ)$, though, this information is not provided to the search algorithm.
The recovered track (with $\text{SNR}_\text{tot}=11$) is a good match.

In Fig.~\ref{fig:inj2}c, we show the spectrogram for an even longer week-long signal present in two weeks of data.
The signal parameters are identical to parameters used in Figs.~\ref{fig:inj2}ab except we use a smaller value of $\dot{f}=\unit[10]{Hz\,day^{-1}}$.
Using $2\times10^6$ cubic B\'ezier templates, we recover the injected signal in Fig.~\ref{fig:inj2}d with $\text{SNR}_\text{tot}=67$.

In Table~\ref{tab:h0}, we show the strain values $h_0$ for which different signals (such as those shown in Fig.~\ref{fig:inj2}) can be detected in $\unit[90]{days}$ of data with a false alarm probability of 1\% and a false dismissal probability of 50\%.
We look for signals in spectrograms of duration $\Delta t=\unit[1]{day}$, and in one case, $\Delta t=\unit[14]{days}$.
We simulate noise to determine the threshold of $\text{SNR}_\text{tot}$ required for 1\% false-alarm probability detection.
Then, we recover simulated signals with varied $h_0$ to determine the value for which at least half of the signals are recovered with a sufficiently large value of $\text{SNR}_\text{tot}$ (false dismissal probability = 50\%).
In all cases, we assume an optimally oriented source with an optimal sky location.
(For randomly oriented sources with random sky locations, we expect $h_0$ to increase by a factor of $\approx$$1.5$--$2$; see, e.g.,~\cite{stochsky}.)
We employ $2\times10^5$ templates in a $\unit[300]{Hz}$ bandwidth.

The results shown in Table~\ref{tab:h0} can be understood in terms of a simple scaling law, which can be used to guide our expectations for the detectability of an arbitrary signal.
If the seedless clustering template approximately matches the signal we are trying to recover, then the detection statistic scales approximately as follows:
\begin{equation}\label{eq:scaling}
  \text{SNR}_\text{tot} \propto \frac{h_0^2}{\overline{P(f)}} 
  \left(\frac{t_\text{dur}}{\Delta f}\right)^{1/2} ,
\end{equation}
where $\overline{P(f)}$ is the strain noise power spectral density averaged over the emission frequency and $t_\text{dur}$ is the duration of emission.
While $t_\text{dur}$, $\Delta f$, and $\overline{P(f)}$ are the most important factors in determining the detectability of a signal, other variables play a role too.
For example, using a fixed number of templates, it is harder to detect signals as the frequency increases (all else equal) because the shrinking diffraction-limited resolution makes it more challenging to match the signal to a template~\cite{stochsky}.

\begin{table}
  \begin{tabular}{|c|c|c|c|c|}
    \hline
    $f_0$ ($\unit[]{Hz}$) & $\dot{f}$ ($\unit[]{Hz\,day^{-1}}$) & $t_\text{dur}$ ($\unit[]{days}$) & $\Delta t$ ($\unit[]{days}$) & $h_0$ \\\hline
    100 & 20 & 0.29 & 1 & $3.6\times10^{-24}$ \\\hline
    100 & 0 & 0.29 & 1 & $3.3\times10^{-24}$ \\\hline
    100 & 10 & 7 & 14 & $1.7\times10^{-24}$ \\\hline
    1000 & 0 & 0.29 & 1 & $6.2\times10^{-24}$ \\\hline
  \end{tabular}
  \caption{
    The detectability of different signals in simulated Advanced LIGO noise.
    The variables $f_0$ and $\dot{f}$ describe the signal morphology following Eq.~\ref{eq:phi} while $t_\text{dur}$ is the signal duration.
    $\Delta t$ is the duration of each spectrogram input to the seedless clustering algorithm.
    The variable $h_0$ is the signal strength that can be recovered with a false alarm probability of 1\% and a false dismissal probability of 50\% in $\unit[90]{days}$ of data.
    We assume optimally oriented sources in optimal sky locations.
    We employ $2\times10^5$ templates per spectrogram and analyze data in a $\unit[300]{Hz}$ bandwidth.
    We use $(\delta t, \Delta f)=(\unit[158]{s},\unit[1]{Hz})$ pixels.
    The sensitivities quoted here can very likely be improved by using more templates.
  }
  \label{tab:h0}
\end{table}


Together, the recoveries demonstrated in Figs.~\ref{fig:days} and~\ref{fig:inj2}, and also Tab.~\ref{tab:h0}, show that it is possible to recover very long-lived signals with a range of different signal morphologies in compressed cross-correlated data.
The recovery works for relatively weak signals and can be carried out with reasonable computational resources.
It is made feasible by coarse-graining, which allows us to analyze days of data using a single spectrogram.

In order to place the results of Tab.~\ref{tab:h0} in context, we consider an illustrative emission scenario.
We consider a neutron star in a binary, which emits gravitational waves, for example, through $r$-mode instabilities, which decay on a timescale of a $\approx$$\unit[1]{week}$.
By assuming accretion-torque balance, it has been hypothesized~\cite{bildsten} that the low-mass X-ray binary Scorpius X-1 emits gravitational waves with a strain of $\approx$$9\times10^{-26}$ at $f=\unit[100]{Hz}$~\cite{sideband_method}.
By tuning the bin width to match the observed Doppler modulation of Scorpius X-1, we expect the sensitivity stated in Tab.~\ref{tab:h0} could be improved by a factor of $(\unit[1]{Hz}/\unit[0.05]{Hz})^{1/4}\approx2$ near $\unit[100]{Hz}$~\footnote{Here we are envisioning a targeted search, which, if necessary, could employ hybrid compression.}.
This is still about a factor of 10 below the expected strain from Scorpius X-1 assuming accretion-torque balance.

Thus, assuming the hypothesized accretion-torque balance scenario, advanced detectors, using the method presented here, are unlikely to detect very long transient gravitational waves from a source like Scorpius X-1 ($d=\unit[2.3]{kpc}$).
Detection would require a source that is either significantly closer or significantly brighter.
That said, it is important to note that detecting {\em any} gravitational waves from neutron stars such as Scorpius X-1 is a challenging proposition for a variety of search techniques~\cite{scox1_mdc}.
The method proposed here will help to cover the full parameter space of possible signals from such sources.

Before we move on, we note in passing, an interesting difference between this very long transient algorithm and the matched filtering algorithm proposed in~\cite{tcw}.
In~\cite{tcw}, the authors note that---at a fixed frequency---the {\em matched filtering} signal-to-noise ratio depends only on the total energy emitted in gravitational waves $E_\text{gw}$ and not on the signal duration $t_\text{dur}$.
However, since radiometric signal-to-noise ratio scales like $\text{SNR} \propto \sqrt{t_\text{dur}} \, h_0^2$, and since $E_\text{gw} \propto t_\text{dur} \, h_0^2$, $\text{SNR} \propto E_\text{gw} / \sqrt{t_\text{dur}}$.
Thus, for radiometer analyses, we obtain a different scaling behavior than the matched filter behavior noted in~\cite{tcw}: given a fixed energy budget, shorter gravitational-wave transients are easier to detect than longer transients.
Given a constant flux $\Phi_\text{gw} \propto dE_\text{gw}/dt$, the signal-to-noise ratio grows with time: $\text{SNR} \propto \Phi_\text{gw} \, \sqrt{t_\text{dur}}$.

\section{Conclusions}\label{conclusions}
We have demonstrated the machinery of an efficient search for very long lived transient gravitational waves.
Such a search will be useful, both looking for very long transient signals from neutron stars, and for understanding the time-dependence of signals identified through persistent radiometer searches and stochastic background searches.
Our procedure relies on coarse-graining to create an intermediate data product that is computationally manageable to work with.
In carrying out our demonstration, we explored the modulations in cross-correlated data due to the rotation of the Earth.
Future work, building on this demonstration, will explore new physics (and instrumental effects) on timescales of hours to days.

We note that the method described in~\cite{tcw}, which builds on existing searches for continuous waves from persistent emitters, can likely achieve better sensitivity than the method proposed here in its domain of utility: sources following the canonical spin evolution typical of an isolated pulsar.
This is because continuous wave searches for isolated neutron stars (see, e.g.,~\cite{cws5_allsky}) assume a heavily constrained parameter space of pulsar-like signals.
However, there are important differences between radiometry and continuous waves searches, which make our new method highly complementary.

Like all applications of gravitational-wave radiometry, the method described here is most useful in cases when the signal does not match a simple template.
For neutron stars in binary systems, radiometer measurements~\cite{sph_results,asaf} can be comparable in sensitivity to continuous wave techniques~\cite{twospect,sideband}, which rely on a more detailed signal model.
Further, radiometer searches, including the one described here, are extremely robust because they do not make assumptions about the phase evolution of the signal.
Finally, in the method described here, we allow for significant variation in the frequency evolution of the gravitational-wave signal, which allows us to probe a completely different parameter space compared to continuous wave searches, which assume very limited changes in the neutron star spin: $\lesssim\unit[6\times10^{-9}]{Hz\,s^{-1}}$ versus $\lesssim\unit[300]{Hz\,s^{-1}}$.

While we have focused here on signals lasting hours to weeks, this method can be extended to target both shorter and longer signals as well.
While significant work has been carried out to develop tools for probing signals with durations of $10$--$\unit[1000]{s}$, it may prove useful to apply some modest coarse-graining ($\Delta f/\delta f \gtrsim 1$) to searches that are stretched for computational resources.
Characterization of $\approx$year-long signals may require some additional development since a year-long spectrogram of compressed data still requires significant memory resources.
However, extrapolating from our analysis here of a two-week-long spectrogram, it seems likely that such a search could be carried out through a number of options such as the use of computers with larger amounts of memory $\gtrsim$$\unit[24]{GB}$ and/or working with narrower frequency bands.

Another next step is to consider how realistic searches, using the ideas presented here, can be optimized.
For example, a search for persistent narrowband signals may benefit from the use of overlapping frequency bins and/or frequency bins of different sizes.
Such optimizations would be equally useful for any radiometer search.

\section{Acknowledgments}
We thank Joe Romano for helpful discussion on coarse-graining, Vladimir Dergachev for discussion on intermittent pulsars, and David Keitel for carefully reading and commenting on an earlier version of the manuscript.
ET was a member of the LIGO Laboratory, supported by funding from United States National Science Foundation.
LIGO was constructed by the California Institute of Technology and Massachusetts Institute of Technology with funding from the National Science Foundation and operates under cooperative agreement PHY-0757058.
VM's work was supported by NSF grant PHY1204944.
NC's work was supported by NSF grant PHY-1204371.
This is LIGO document P1400256.

\begin{appendix}
\section{Details of radiometry}\label{details}
\subsection{Definitions}
Given two simultaneous strain measurements $s_1$ and $s_2$, the cross-correlation statistic $\widehat{Y}$ (and its associated uncertainty $\sigma$) can be written as~\cite{stamp}:
\begin{equation}\label{eq:Y}
  \begin{split}
  \widehat{Y}(t;f|\hat\Omega) & = \frac{2}{\cal N} \text{Re}\left[ 
    e^{2\pi i f \hat\Omega \cdot \Delta \vec{x}/c}
    \frac{1}{\epsilon_{12}(t|\hat\Omega)}
    \tilde{s}_1^*(t;f) \tilde{s}_2(t;f)
    \right] \\
  \sigma^2(t;f|\hat\Omega) & = \frac{1}{2}
  \left|\frac{1}{\epsilon_{12}(t|\hat\Omega)}\right|^2 P'_1(t;f) P'_2(t;f) .
  \end{split}
\end{equation}
Here, $\tilde{s}_I(t;f)$ is the discrete Fourier transform of the strain series from detector $I$ and $P'(t;f)$ is the auto-power spectrum calculated using neighboring data segments.
Each segment has duration $\delta t$.
The argument $t$ is the starting time of the data segment and $f$ is frequency.
${\cal N}$ is a Fourier normalization coefficient, $\hat\Omega$ is the direction of the source, $\Delta\vec{x}$ is the difference in detector position, and $c$ is the speed of light.
Above, we carry out simulations for the Advanced LIGO network (consisting of detectors in Hanford, WA and Livingston, LA) for which $\left|\Delta\vec{x}\right|/c\approx\unit[10]{ms}$.
The $\epsilon_{12}(t|\hat\Omega)$ factor describes the efficiency of the detector pair~\cite{stamp}:
\begin{equation}\label{eq:epsilon}
  \epsilon_{12}(t|\hat\Omega) \equiv \frac{1}{2} 
  \sum_A F_1^A(t|\hat\Omega) F_2^A(t|\hat\Omega) .
\end{equation}
Here, $F_I^A(t|\hat\Omega)$ is the antenna factor~\cite{300years} for detector $I$ and $A=+,\times$ are the different polarization states.

It is useful to define ``pixel signal-to-noise ratio'':
\begin{equation}\label{eq:rho}
  \rho(t;f|\hat\Omega) = \widehat{Y}(t;f|\widehat\Omega) \Big/ 
  \sigma(t;f|\hat\Omega) .
\end{equation}
We use $\rho(t;f|\hat\Omega)$ for its convenience as a tool for visual representation of cross-correlated data and also because it allows for compact derivations.
In the context of all-sky searches, it is also useful to define ``complex signal-to-noise ratio''~\cite{stochsky}
\begin{equation}\label{eq:complex}
  \mathfrak{p}(t;f) = \kappa\,
  \tilde{s}_1^*(t;f) \tilde{s}_2(t;f) \Big/ \sqrt{P'_1(t;f) P'_2(t;f)} ,
\end{equation}
where, for the sake of compactness, we define a normalization factor
\begin{equation}\label{eq:kappa}
  \kappa \equiv 2\sqrt{2}/{\cal N} .
\end{equation}
Comparing Eq.~\ref{eq:rho} and~\ref{eq:complex}, we see that $\mathfrak{p}(t;f)$ is the same as $\rho(t;f)$ except no phase factor has been applied, and we do not take the real part.
In doing so, we obtain a data product with information from all directions on the sky.

\subsection{Clustering}
Having defined spectrograms of $\rho$ or $\mathfrak{p}$ (Eqs.~\ref{eq:rho} and~\ref{eq:complex}), the next step in a radiometer analysis is to look for clusters of statistically significant excess coherence.
Details of the clustering procedure depend on the specific search, but there are certain common features.

First, a rule is defined for how clusters can be formed.
In the case of the targeted narrowband radiometer~\cite{radiometer,radio_method,sph_results}, the signal is assumed to persist with a fixed frequency, and so each cluster corresponds to the set of all pixels at a given frequency; i.e., one spectrogram row.

Once a cluster is defined, all the pixels in that cluster are combined in order to produce a detection statistic for that cluster denoted $\text{SNR}_\text{tot}$.
The detection statistic (for a targeted search with a known sky location) can be written as as a weighted average~\cite{radiometer,radio_method,sph_results}:
\begin{equation}\label{eq:snr}
  \text{SNR}_\text{tot} = \frac{\sum_{i\in\Gamma} \widehat{Y}_i \, 
    \sigma_i^{-2}}
       {\left(\sum_{i\in\Gamma} \sigma_i^{-2}\right)^{1/2}} .
\end{equation}
(We have suppressed the dependence on $\hat\Omega$ for the sake of brevity.)
The sum over $i$ runs over some set of $(t;f)$ pixels $\Gamma$, which is determined by details of the search.
A search for persistent narrowband signals, for example, might sum over time at a fixed frequency.
In searches for transient signals, the set of $(t;f)$ pixels may be described, e.g., by a B\'ezier curve~\cite{stochtrack,stochsky,stochtrack_cbc}.

If we assume for the sake of simplicity that the noise is approximately stationary, and that the signal does not vary significantly in frequency, then we can write
\begin{equation}\label{eq:rho2}
  \begin{split}
    \widehat{Y}_i & = \rho_i \, \sigma_0 / \epsilon^i_{12} \\
    \sigma_i & = \sigma_0 / \epsilon^i_{12} .
  \end{split}
\end{equation}
Recall that $\epsilon^i_{12}$ (Eq.~\ref{eq:epsilon}) is an efficiency factor arising from the detector response function, which changes in time as the Earth rotates.
The variable $\sigma_0$ is some number determined only by the detector strain sensitivities.
Combining Eqs.~\ref{eq:snr} and~\ref{eq:rho2}, we obtain:
\begin{equation}\label{eq:vlong}
  \text{SNR}_\text{tot} = \frac{\sum_{i\in\Gamma} \rho_i \, \epsilon^i_{12}}
       {\left(\sum_{i\in\Gamma} (\epsilon^i_{12})^2 \right)^{1/2}} .
\end{equation}

\subsection{All-sky search}
To this point in the appendix, we have assumed that the sky location $\hat\Omega$ is known a priori.
This assumption holds for targeted searches, e.g., searching for gravitational waves from the low-mass X-ray binary, Scorpius X-1~\cite{bildsten}.
When the sky location is not known, the search becomes more complicated.
To see this, consider an all-sky search which considers some finite set of directions $\left\{\hat\Omega_i\right\}$.
If we look in some direction $\hat\Omega_i$, which is offset from the true source direction $\hat\Omega_t$, then the phase factor in Eq.~\ref{eq:Y} will not rotate the signal power perfectly into the real number axis.
As a result, signal power leaks into the imaginary direction, and can even produce negative power on the real axis.

In order to carry out an efficient all-sky search, we can use Eqs.~\ref{eq:vlong} and~\ref{eq:complex} to define a detection statistic that can quickly scan over sky locations:
\begin{equation}\label{eq:vlong_as}
  \text{SNR}_\text{tot}(\hat\Omega) = \frac{\text{Re}\left[
    \sum_{i\in\Gamma}
    \exp\left(-i \Psi_i\right)
    \mathfrak{p}_i \, \epsilon_{12}^i
    \right]
  }{
    \left(\sum_{i\in\Gamma}(\epsilon_{12}^i)^2\right)^{1/2}
  } .
\end{equation}
(This definition of $\text{SNR}_\text{tot}$ is essentially the same one that is proposed in~\cite{stochsky}.)
Graphically, Eq.~\ref{eq:vlong_as} can be interpreted as applying an $\hat\Omega$-dependent phase array to a complex-valued spectrogram (as in Figs.~\ref{fig:days}a and~\ref{fig:days}b) before summing over clusters of pixels.

At first glance, Eq.~\ref{eq:vlong_as} may appear tautological.
There is already implicit dependence on $\hat\Omega$ in Eq.~\ref{eq:vlong} hiding in $\rho_i$, $\epsilon_i$, and therefore in $\text{SNR}_\text{tot}$.
However, the ingredients used in Eq.~\ref{eq:vlong} are calculated for a specific sky location $\hat\Omega$ in such a way that the information about all other sky locations is lost when the imaginary part is discarded in Eq.~\ref{eq:Y}.
Thus, the formulation described by Eq.~\ref{eq:vlong_as} allows us to search the entire sky with a single data product.
Using parallel computing architecture, we can efficiently search many directions in parallel with a single manageable array, dramatically speeding up calculations in the process~\cite{stochsky}.
\end{appendix}

\bibliography{verylong}

\end{document}